\documentclass[showpacs,twocolumn,superscriptaddress]{revtex4-1}
\usepackage{color, amsmath, bm}
\usepackage{bbm}
\usepackage{graphicx}
\usepackage{amssymb}
\usepackage{dirtytalk}
\usepackage{natbib}
\usepackage{physics} 
\usepackage{float}
\newcommand*{\citen}{}
\DeclareRobustCommand*{\citen}[1]{%
 \begingroup
 \romannumeral-`\x 
 \setcitestyle{numbers}%
 \cite{#1}%
 \endgroup
}

\begin{document}

\title{Detectable Signature of Quantum Friction on a Sliding Particle in Vacuum}

\author{Fernando C. Lombardo}
\affiliation{Departamento de Física Juan José Giambiagi, FCEyN-UBA. Ciudad Universitaria, Pabellón I, 1428 Buenos Aires, Argentina.}
\affiliation{IFBA CONICET-UBA.Ciudad Universitaria, Pabellón I, 1428 Buenos Aires, Argentina.}
\author{Ricardo S. Decca}
\affiliation{Department of Physics, Indiana University-Purdue University Indianapolis.
 Indianapolis, Indiana 46202, USA}
\author{Ludmila Viotti}
\affiliation{Departamento de Física Juan José Giambiagi, FCEyN-UBA. Ciudad Universitaria, Pabellón I, 1428 Buenos Aires, Argentina.}
\author{Paula I. Villar}
\affiliation{Departamento de Física Juan José Giambiagi, FCEyN-UBA. Ciudad Universitaria, Pabellón I, 1428 Buenos Aires, Argentina.}
\affiliation{IFBA CONICET-UBA.Ciudad Universitaria, Pabellón I, 1428 Buenos Aires, Argentina.}



\begin{abstract}
Spatially separated bodies in relative motion through vacuum experience a tiny friction force known as quantum friction. This force has so far eluded experimental detection due to its small magnitude and short range. Quantitative details revealing traces of the quantum friction in the degradation of the  quantum coherence of a particle are presented. Environmentally induced decoherence for a particle sliding over a dielectric sheet can be decomposed into contributions of different signatures: one solely induced by the electromagnetic vacuum in presence of the dielectric and another induced by motion. 
As the geometric phase has been proved to be a fruitful venue of investigation to infer features of the quantum systems,  herein we propose to use the accumulated geometric phase acquired by a particle as a quantum friction sensor. Furthermore, an innovative experiment designed to track traces of quantum friction by measuring the velocity dependence of corrections to the geometric phase and coherence
is  proposed. The experimentally viable scheme presented can spark renewed optimism for the detection of non-contact friction, with the hope that this non-equilibrium phenomenon can be readily measured soon.
\end{abstract}

\maketitle

\section{Introduction}

One of the most exciting characteristics of quantum field theory the non-trivial structure of the vacuum state.
The idea that the quantum vacuum is a seething sea of fluctuations can be traced back to the early decades of quantum physics. Due to the Heisenberg principle, energy and time constitute a pair of conjugated variables: the more you know about one of them, the less you know about the other. It then follows that the zero-energy of the vacuum can never be precisely known. Particularly, the uncertainty principle applied to the electromagnetic field implies that it can never be turned off since the fields $\mathbf E$ and $\mathbf B$ do not simultaneously commute. According to quantum theory, the quantum vacuum is filled by short-lived particles, continuously appearing and disappearing, filling the vacuum with a non-zero \say{zero-point energy}. Much controversy has been raised around this zero-point energy. Wolfgang Pauli's words in his 1945 Nobel lecture were \say{It is clear that this zero-point energy has no physical reality}. Luckily, those words did not discourage further investigations at that moment. Evidence began to accumulate indicating that this lively vacuum had macroscopic observable consequences such as the modification of the energy between the levels of an atom, known as the Lamb shift. 
In 1948, Hendrick Casimir\cite{Casimir} showed that the emergence of an attractive force between two uncharged, perfectly conducting parallel plates, the so-called Casimir effect, was due to quantum vacuum fluctuations. Another fascinating effect occurs when a mirror moves through space at relativistic speeds: some photons become separated from their partners and the mirror begins to produce light. This phenomenon is known as Dynamical Casimir Effect (DCE) \cite{DCE}.

\vspace{0.3cm}
Some outstanding features of modern quantum field theory are due to the non-trivial structure exhibited by the vacuum state and the consequent existence of vacuum fluctuations. These quantum fluctuations induce macroscopic effects for which, in many cases, experimental verification has been achieved and thereafter improved. So far, the static Casimir force has been measured \cite{Casimir,book_milonni, Lamoreaux1997, bordag1, bordag2, book_milton, milton2004casimir, reynaud2001quantum}. However, the challenge involved in making a mirror (or a neutral particle) move at an almost relativistic speed in a medium has prevented direct observation of most dynamical effects. The only macroscopic effect that has allowed experimental detection of the dynamical Casimir effect was based on electromagnetic analogs of a moving mirror using a tunable reflecting element in a superconducting circuit (or cavities with time-dependent electromagnetic properties) \cite{review_friction,nation_colloquium,dyncasexp_supercond,dyncasexp_squid}.
But while experimental observation has been attained on Casimir static and dynamic effects, a lesser known phenomenon called quantum friction (QF) \cite{pendry97,pendry_debate,Pendry_reply, Leonhardt_2010, Leonhardt_nofriction, volokitin_persson} has eluded detection so far. This is mainly due to QF's short range and small magnitude. The lack of verification has led to co-existence of different theoretical approaches that rely on a variety of assumptions and do not converge to a single result. The various methods ranges from time-dependent perturbation theory \cite{barton_atom_halfspace, intravaia_acceleration}, quantum master equations in the Markovian limit \cite{Scheel}, to generalized non-equilibrium fluctuation-dissipation relations \cite{dalvit_fluctuation} and thermodynamic principles \cite{hu_thermodynamic}. 

\vspace{0.3cm}
The emergence of different theoretical approaches contributing to controversy among the Casimir physics community can only be eased by the experimental detection of such a force. Due to the experimental challenges involved in the implementation of precision measurements for the observation of such a small force acting on objects near a surface, there has been lately a set of works devoted to finding favorable conditions for its detection
\cite{farias2018quantum, carusotto2017friction,intravaia_rolling,farias_graphene,farias_friction,viotti_thermal,lufriction2}. 
In Refs. \citen{volokitin2011quantum, volokitin_cherenkov}, authors have investigated the van der Waals friction between graphene and an amorphous SiO2 substrate. They calculated the frictional dragging between two graphene sheets caused by van der Waals friction and proved that this dragging can induce a voltage large enough to be measured experimentally by state-of-the-art non-contact force microscopy. This work paved the way for possible mechanical detection of the Casimir friction. In \citen{buhmann_spectroscopic}, the modifications arising from the motion in the level shift and decay rate of an atom in the presence of a medium are found in the Markovian limit, and their relation to the parallel component of the vacuum force is discussed. 
Recently, in \citen{farias_nature} some of us have examined the effect of the vacuum, dressed by the presence of a more realistic Drude-Lorentz material on the geometric phase acquired by an atom moving at constant velocity and made a proposal for an experimental setup. 

\vspace{0.3cm}
A natural question arises at this stage: can we track traces of quantum friction? Considering how difficult it is through direct measurement of the force, maybe its challenging detection should be pursued by a different approach. Hence, in this manuscript, we present a novel proposal. The idea borrows a fundamental brick of quantum computation and uses it as a sensor of a particle's degree of freedom: the phase-lag induced by the motion on the fluctuating dipole results in effects which are illustrated in Section \ref{QF}. Section \ref{model} explains the model studied and Section \ref{environment} exhibits the evidence of the motion-induced environmental effects on the quantum coherence of the particle. These corrections due to the quantum fluctuations can be seen as stochastic variations in the energy gap of the two-level system. Section \ref{GP} explores how these variations on the internal degree of freedom of the particle are manifested in corrections of the accumulated geometric phase acquired in a non-unitary evolution. The mere presence of a velocity contribution in the noise corrections to the geometric phase is an indirect manifestation of the quantum friction force.
The experimental proposal in Section \ref{experiment} puts forward a state-of-the-art detection scheme and considers realistic values of the parameters involved. 
While it may be argued that the controversy on a non-contact friction force only enthuses theoretical researchers due to its tiny magnitude  (i.e.) understanding QF will not readily help build better turbines, road surfaces, or cars ), the many emerging micro and nano-mechanical systems that promise new applications in sensors or information technology may suffer - or even benefit- from non-contact friction. In such cases, a better understanding of QF is only the first step, and hopefully our work can inpire renewed optimism in the design of new experimental setups for the detection of non-contact friction with the goal that this phenomenon can be measured in the near future.

\vspace{0.3cm}
\section{Bodies in relative motion: quantum friction}
\label{QF}

\begin{figure}[h]
\centering
\includegraphics[ width=.75\columnwidth]{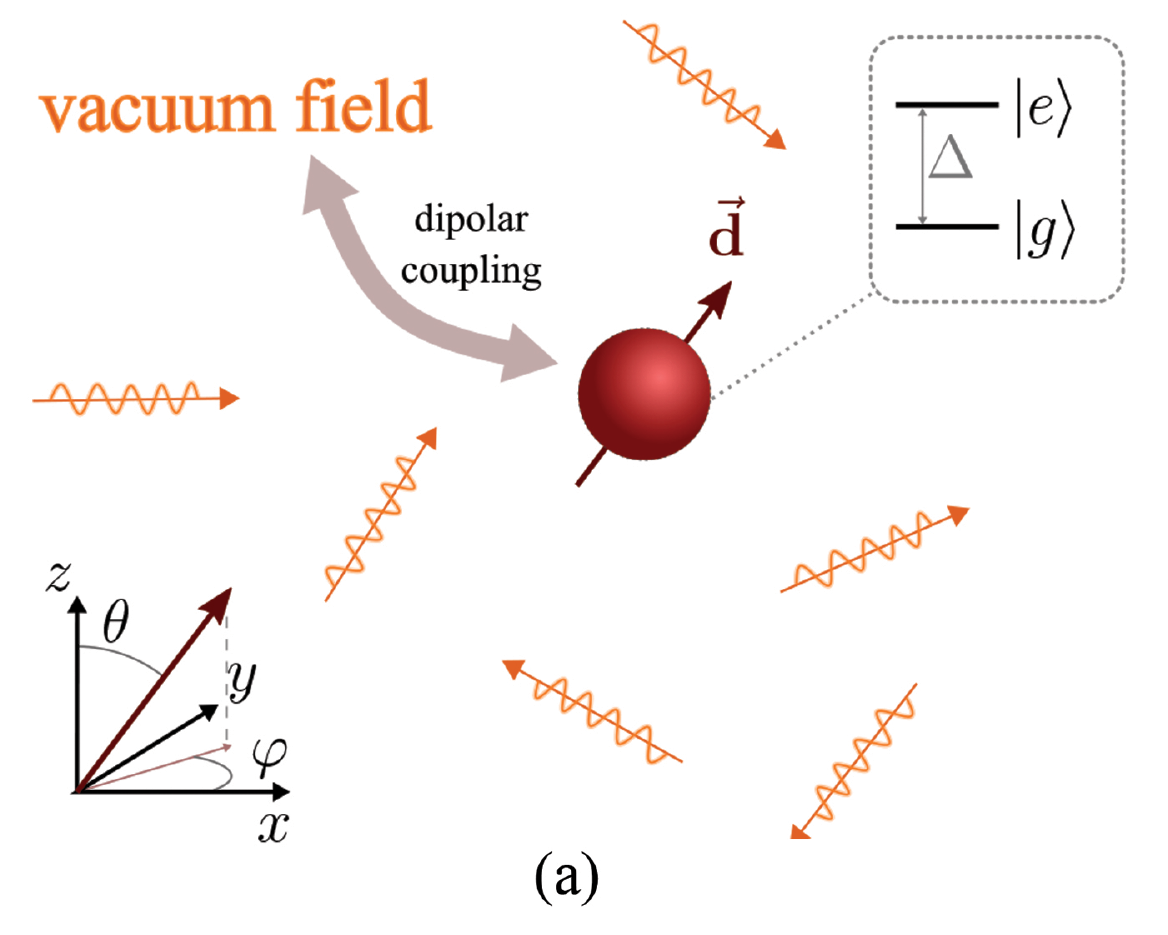}
\includegraphics[ width=.75\columnwidth]{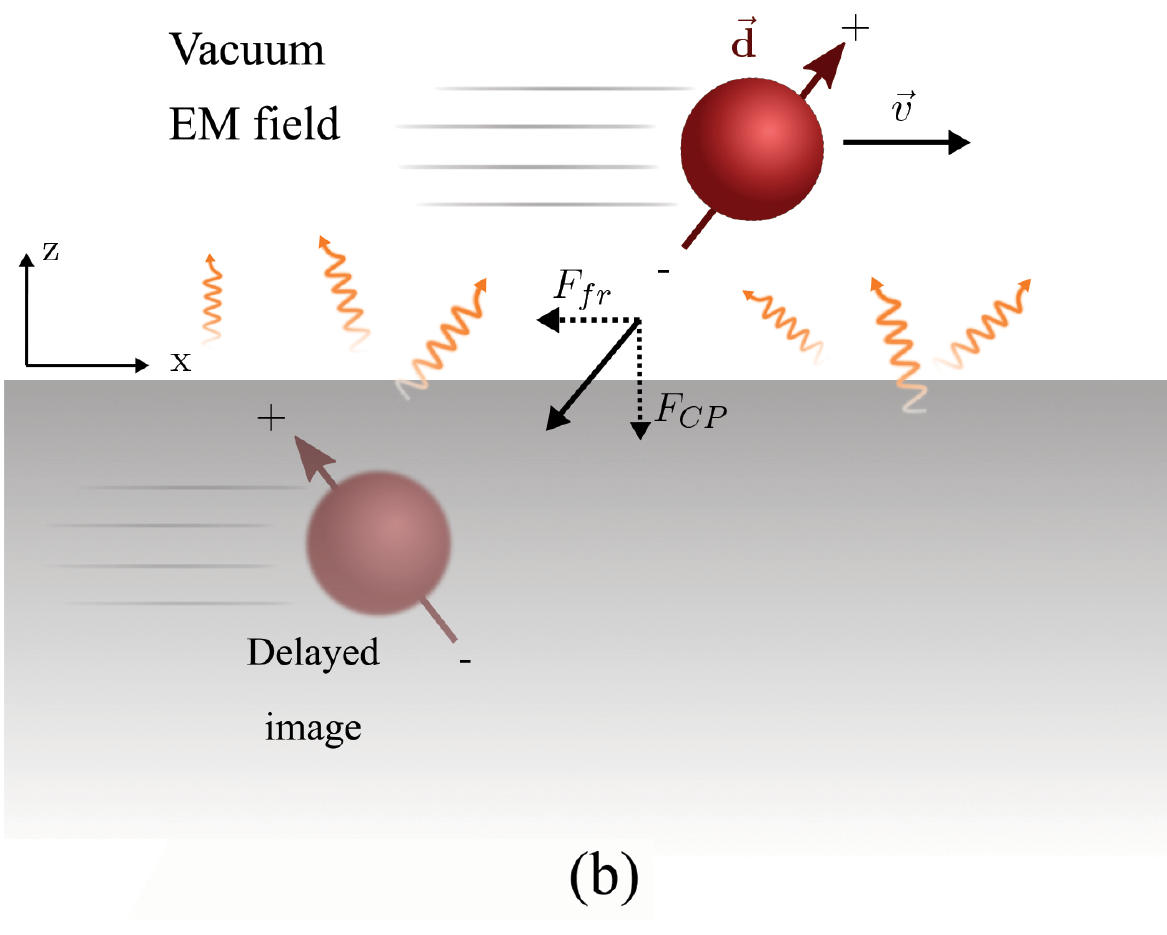}
\caption{\label{fig1}(a)  Emergence of a fluctuating dipole moment in a neutral particle embedded in an electromagnetic field. (b) Emergence of a quantum friction force $F_{fr}$ acting on the particle when it moves over a non-perfect surface.}
\end{figure}

We start with an introductory description of the curious phenomenon under study. Assuming there is a neutral particle embedded in an electromagnetic field at zero temperature, the vacuum quantum fluctuations will induce a fluctuating dipole moment in the particle. If this system is brought close to a neutral conducting surface, the response of the macroscopic material can be described by an image fluctuating dipole at the mirrored position of the real particle. Thus, as the dipole oscillates, the image dipole oscillates as well.  These oscillations result in  an attractive force pulling the neutral particle towards the surface. This fluctuation-induced force is called the van der Waals or Casimir-Polder force. If the particle is now set in constant relative motion parallel to the surface, the mirrored dipole will attempt to follow. As long as the surface is a perfect conductor, the situation is equivalent to the static case since the charges in the macroscopic surface re-accommodate immediately. However, if the particle travels over a dielectric surface, the image dipole will lag due to the resistivity of the material. Real and image dipole oscillations' generate a tilted force.  The additional parallel component opposed to the motion's direction is the Casimir or quantum friction force. The phenomenon is illustrated in \textbf{Figure \ref{fig1}}. 

\subsection{Model}
\label{model}

\begin{figure}
\centering
 \includegraphics[width=0.9\columnwidth]{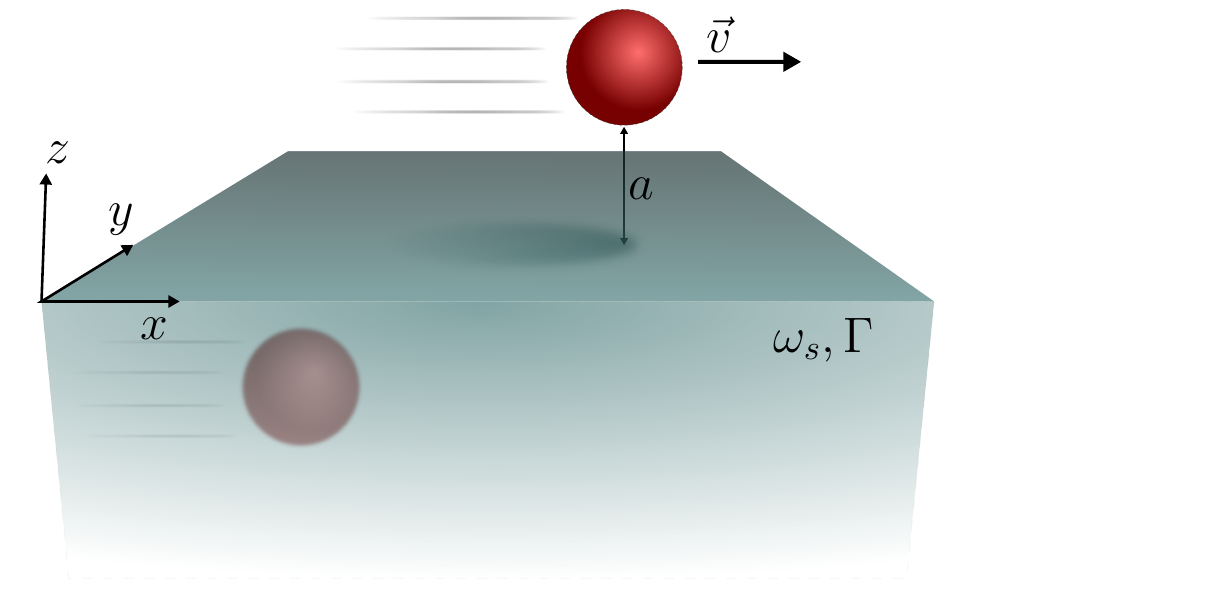}
 \caption{Schematic model: a two-level atom moving with constant velocity $\vec{v}$ at a fixed distance $a$ from a  dielectric plate.}
 \label{fig2}
\end{figure}

\vspace{0.3cm}
We shall consider a neutral particle moving through a medium-assisted electromagnetic field vacuum. As shown in \textbf{Figure \ref{fig2}}, the particle is modeled as a two-level system for which the center of mass follows a prescribed trajectory $\mathbf{r}_{\text{s}}(t) = v\,t\;\hat{x} + a\;\hat{z}$ at a fixed distance $a$ from a dielectric semi-infinite planar medium. The dynamic of the composite system can be described by a Hamiltonian consisting of a system, field, and interaction contributions defined by

\begin{equation}
 \hat{H}=\frac{\hbar}{2}\Delta\; \hat\sigma_\text{z} \otimes \mathbf{1} + \hat{H}_{\text{em}}+\hat{H}_{\text{int}},
\end{equation}
where $\Delta$ is the energy gap of the two-level system and $\hat{H}_{\text{em}}$ is the Hamiltonian of the electromagnetic field in absence of the particle but in presence of the dielectric half-space $z<0$. The interaction between the particle and the field is given in the dipole (long wavelength) approximation by $\hat{H}_{\text{int}} = - \hat{\mathbf{d}}\,\otimes\, \hat{\mathbf{E}}(\mathbf{r}_{\text{s}})$ and depends explicitly on time through the position of the particle, which is treated as a classical variable since its momentum-position uncertainty is unresolvable by the characteristic wavelength of the electric field.
We shall restrict ourselves to the non-retarded (near field) regime where the particle-surface distance $a$ is small enough to satisfy $a \Delta/c\ll1$. In this regime, the finite time taken for a reflected photon to reach the particle is negligible compared to its natural timescale and the interaction Hamiltonian can therefore be written as $\hat{H}_{\text{int}} = \hat{\mathbf{d}}\otimes \nabla\hat{\Phi}(\mathbf{r}_{\text{s}})$. The electric potential $\hat{\Phi}$, expanded in a plane-wave basis corresponding to elementary excitations, is \cite{barton78,barton79}

\begin{equation}
 \hat{\Phi}=\int d^2k\;\int_0^\infty d\omega \left(\hat{a}_{\mathbf{k},\omega},\phi(\mathbf{k},\omega)e^{i \mathbf{k} \mathbf{r}_\parallel} + h.c.\right),
\end{equation}
and contains all the information of the electric field in the $z>0$ region, dressed by the dielectric medium. The bosonic operators satisfy the commutation relation $\left[\hat{a}_{\mathbf{k},\omega},\hat{a}^\dagger_{\mathbf{k}',\omega'}\right]=\delta\left(\mathbf{k}-\mathbf{k}'\right)\delta\left(\omega-\omega'\right)$, and the single excitation mode functions are given by

\begin{equation}
 \phi\left(\mathbf{k}, \omega\right)=\sqrt{\frac{\omega\Gamma}{s}}\sqrt{\frac{\hbar}{2\pi^2 k}}e^{-k\,z}\frac{\omega_{\text{p}}}{\omega^2-\omega_{\text{s}}^2 - i \omega \Gamma},
\end{equation}
where the wave vector $\mathbf{k}=(k_x, k_y)$ is parallel to the medium surface and $k=|\mathbf{k}|$. The frequency $\omega_{\text{s}}$ gives the surface plasmon resonance and the material dissipation rate $\Gamma$ its broadening and , in the Drude model for metals, the plasma frequency $\omega_{\text{p}}$ satisfies $\omega_{\text{p}}^2= 2\omega_{\text{s}}^2$.

\vspace{.2cm}

With this complete set of mathematical tools, we shall study how the degrees of freedom of the neutral particle are modified due to non-unitary effects induced by the presence of the quantum vacuum and the dielectric surface. Furthermore, we shall get an insight into the crucial role of the velocity since the mere presence of a motion-induced contribution in the environmentally-induced effects is an indication of the frictional effect over the quantum degrees of freedom of the particle.

\subsection{Environmentally motion-induced decoherence}
\label{environment}

In the following, we shall focus on the dynamics of the quantum system. Thus, we must obtain the master equation for the reduced density matrix of the quantum system derived by integrating out the degrees of freedom of the composite environment. By assuming an initially factorized state $\rho(0) = \rho_{\text{s}}(0)\,\otimes\,\rho_{\text{em}}^{\text{vac}}$ with the dressed electromagnetic field in its vacuum state, the master equation in the Schrodinger picture, up to second order in the coupling constant is given by

\begin{align}\nonumber
 \dot{\rho}_{\text{s}}=&-\frac{\mathfrak{i}\Delta}{2}\left[\hat{\sigma}_z,\rho_{\text{s}}\right]-D(v,t)\left[\sigma_x,[\sigma_x,\rho_{\text{s}}]\right]\\[.75em]
&-f(v,t)\left[\sigma_x,[\sigma_y,\rho_{\text{s}}]\right]+\mathfrak{i}\zeta(v,t)\left[\sigma_x,\{\sigma_y,\rho_{\text{s}}\}\right],
 \label{meq_exacta}
\end{align}
where we will consider the particular initial state $\ket{\psi(0)}= \cos(\alpha)\ket{g}+\sin(\alpha)\ket{e}$ throughout this work (we shall assume $\alpha = \pi/4$ for all plots).
The non-unitary effects are modeled by the diffusion coefficients $D(v, t)$ and $f(v, t)$, while dissipative effects are present in $\zeta(v, t)$. All three coefficients are real functions of time, with parameters introduced by the particle and the medium assisted field:

\begin{widetext}
\begin{align}
 D(v,t)&= \frac{r_0}{2\pi}\int_0^t dt'\;\int_0^\infty d\omega \frac{\tilde{\Gamma}\; \omega}{\left(\omega^2-1\right)^2+\tilde{\Gamma}^2\omega^2} \cos(\tilde{\Delta}t')\cos(\omega t') \mathbf{P}(ut') \label{D_raw}\\
 f(v,t)&= \frac{r_0}{2\pi}\int_0^t dt'\;\int_0^\infty d\omega \frac{\tilde{\Gamma}\; \omega}{\left(\omega^2-1\right)^2+\tilde{\Gamma}^2\omega^2} \sin(\tilde{\Delta}t')\cos(\omega t') \mathbf{P}(ut')\label{f_raw}\\
 \zeta(v,t)&= \frac{r_0}{2\pi}\int_0^t dt'\;\int_0^\infty d\omega \frac{\tilde{\Gamma}\; \omega}{\left(\omega^2-1\right)^2+\tilde{\Gamma}^2\omega^2} \sin(\tilde{\Delta}t')\sin(\omega t') \mathbf{P}(ut'),\label{Z_raw}
\end{align}
\end{widetext}

where we have defined a dimensional coefficient $r_0 = (d^2\; \omega_{\text{p}}^2)/(\hbar\;\omega_{\text{s}}^2\;a^3)$ and the dimensionless parameters 
\begin{equation}
 u=\frac{v}{\omega_{\text{s}}\times a}\quad ;\quad \Tilde{\Delta}=\frac{\Delta}{\omega_{\text{s}}}\quad ;\quad \Tilde{\Gamma}=\frac{\Gamma}{\omega_{\text{s}}},
\end{equation}
and changed to dimensionless variables $\omega/\omega_{\text{s}} \rightarrow \omega$ and $t\;\omega_{\text{s}}\rightarrow t$.
$\mathbf{P}(ut')$ is an algebraic function given by

\begin{equation}
 \mathbf{P}(ut') = \frac{2\text{n}_x^2(2 -u^2 t'^2)}{\left(4 +u^2 t'^2\right)^{5/2}} \nonumber + \frac{\text{n}_y^2 }{\left(4 +u^2 t'^2\right)^{3/2}}  + \frac{ \text{n}_z^2 \left(8-u^2 t'^2\right)}{\left(4 +u^2 t'^2\right)^{5/2}}.
\end{equation}
(details can be found in \citen{lufriction2}).

\begin{figure}
\centering
 \includegraphics[width=0.9\columnwidth]{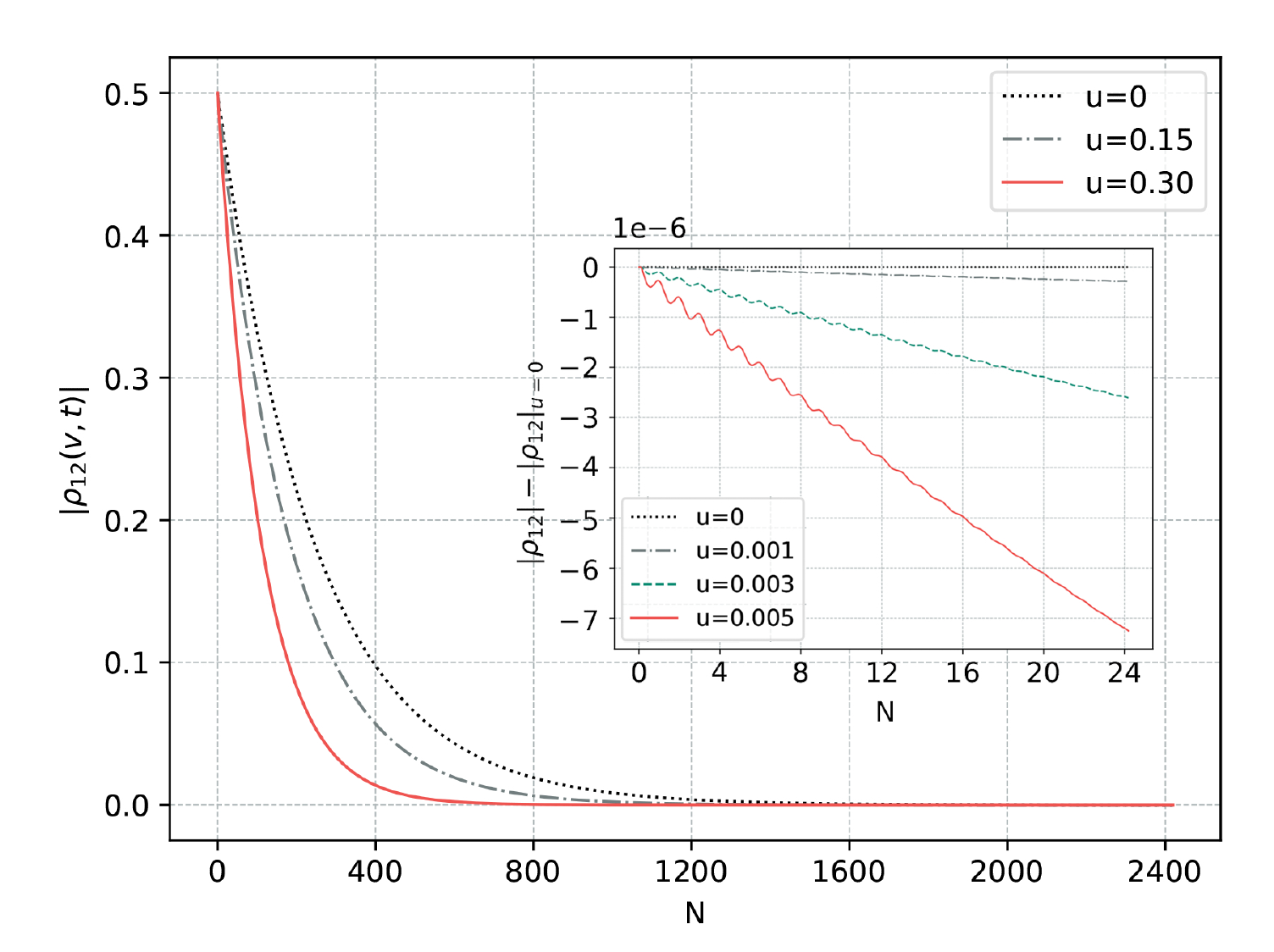}
 \caption{Coherence evolution in natural cycles $N = t/(2\pi/\tilde{\Delta})$ for different velocity values. The insert shows the difference in the coherence as cycles proceed for smaller velocity values. The polarization direction is parallel to the velocity and parameter values are $\tilde{\Gamma}=1$, $r_0/\omega_{\text{s}}=10^{-2}$, $\tilde{\Delta}=0.2$}
 \label{fig3}
\end{figure}

\vspace{0.3cm}
In \textbf{Figure \ref{fig3}} we show the off-diagonal elements of the density matrix, which are suppressed by the environment. This destruction is not only seen to be sped up by the relative motion between the particle and the material but also shows an explicit dependence on the velocity; it occurs sooner as the velocity is increased. The same monotonic behavior is displayed in the insert, in which the difference between absolute values $|\rho_{12}(t)|-|\rho_{12}^{u=0}(t)|$ grows faster as the velocity increases.

\vspace{0.3cm}
The density matrix obtained in Equation \ref{meq_exacta}, under the secular approximation (also known as post-trace rotating wave approximation), allows the definition of a decoherence timescale $\tau_{\text{D}}$.
The decoherence function is known to be ${\cal D}(t) = e^{-\frac{ 2}{\omega_{\text{s}}}\int_0^t dt'\;D(v,t')}$. Hence, we shall set the decoherence timescale as 
 ${\cal D}(\tau_{\text{D}}) = e^{-2}$. It is important to mention that the diagonal elements of the reduced density matrix are exactly the same whether the approximation is performed or not since it only implies disregarding a dynamical interaction between $\rho_{12}$ and $\rho_{21}$ for this system.
In the Markovian limit, and to second order in the dimensionless velocity $u$ which we take to be small, the decoherence timescale behaves as

\begin{equation}
 \tau_{\rm D} = \frac{\hbar \omega_{\text{s}}^2\,a^3}{d^2\omega_{\text{p}}^2}\frac{32}{d^{(i)}}\left(\frac{1}{h(\tilde{\Delta},\tilde{\Gamma})}-\frac{3}{8}\frac{d^{(a)}}{d^{(i)}}u^2\frac{\partial_{\tilde{\Delta}}^2h(\tilde{\Delta},\tilde{\Gamma})}{h^2(\tilde{\Delta},\tilde{\Gamma})}\right)
 \label{tdec_mark},
\end{equation}

where the dependence on the polarization orientation is encoded in $d^{(i)} = 1 + n_z^2$ and $d^{(a)} = 3n_x^2 + n_y^2 + 4n_z^2$, with $\mathbf{d}= d(n_x \hat{x} + n_y \hat{y} + n_z\hat{z})$. The functions $h({\tilde\Delta},{\tilde\Gamma})$ and $g({\tilde\Delta},{\tilde\Gamma})$ are defined by 
\begin{align}\nonumber
    h(\tilde{\Delta},\tilde{\Gamma}) &= \frac{\tilde{\Delta}\tilde{\Gamma}}{(\tilde{\Delta}^2-1)^2+\tilde{\Delta}\tilde{\Gamma}}\\[.75em]\nonumber
    g(\tilde{\Delta},\tilde{\Gamma}) &=\Re\left[\left(1+\frac{2\mathfrak{i}}{\pi}\log(\tilde{\omega}_{\text{r}}/\tilde{\Delta})\right)\right.\\
    &\hspace{1cm}\left.\left(\frac{1}{(\tilde{\omega}_{\text{r}}+\tilde{\Delta})^2}+\frac{1}{(\tilde{\omega}_{\text{r}}-\tilde{\Delta})^2}\right)\right]
\end{align}
with ${\tilde\omega}_{\text{r}}= 1/\sqrt{2} \, \sqrt{2- {\tilde\Gamma} + \mathfrak{i}\sqrt{4-{\tilde\Gamma}}}$. 

\begin{figure}
\centering
 \includegraphics[width=0.9\columnwidth]{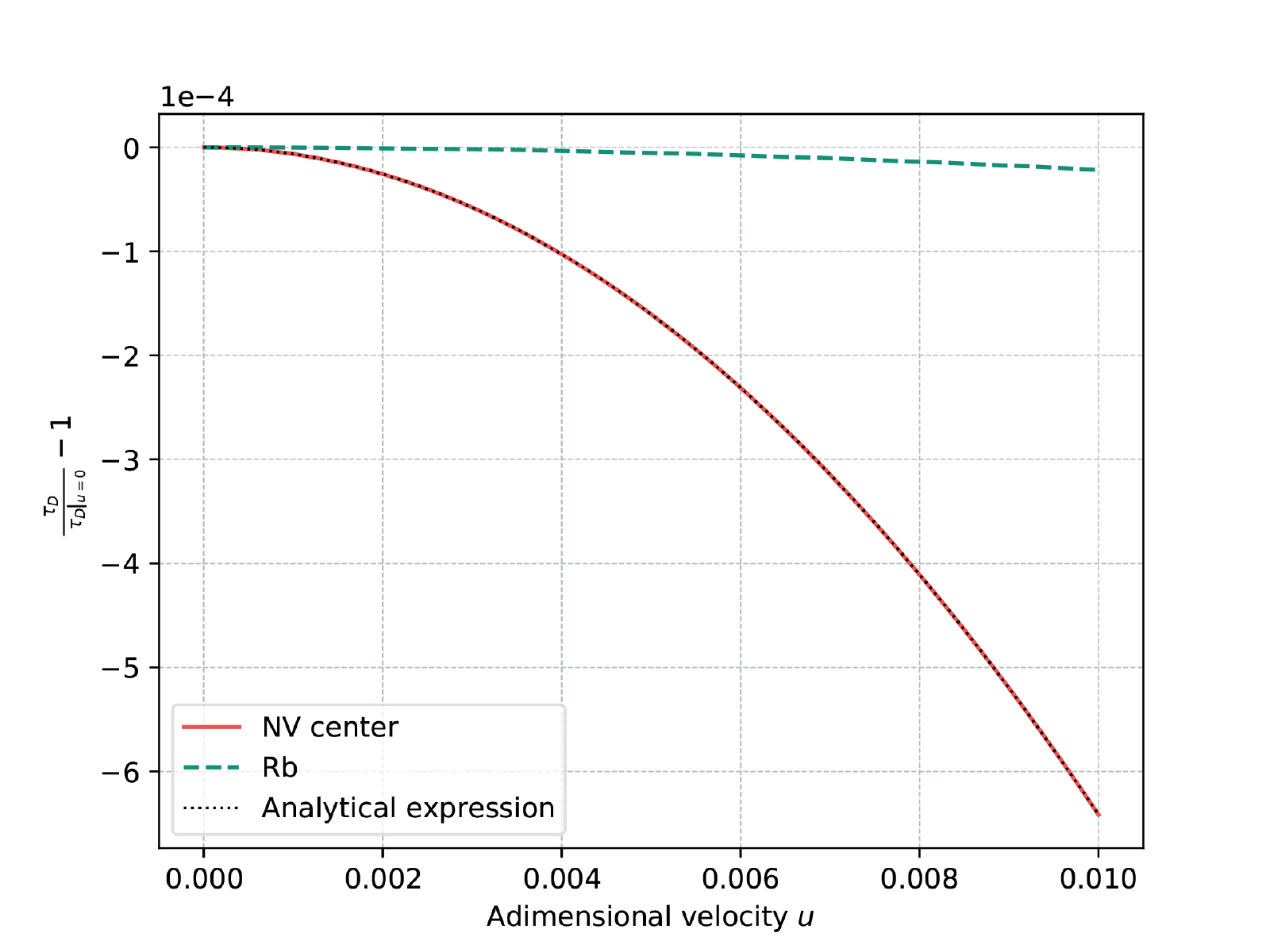}
 \caption{\label{fig4}Decoherence time rate as a function of the adimensional velocity $u$ for dipole orientation parallel to motion. Parameters' values are $\tilde{\Gamma}=1$, $\tilde{\Delta}^{NV}=0.2$/$\tilde{\Delta}^{Rb}=8$, and $r_0/\omega_\text{s}=10^{-2}$.}
\end{figure}

\vspace{0.3cm}
The dependence of the dynamics upon the velocity of the particle can be studied by the use of the decoherence time $\tau_{\text{\rm D}}$, which happens to scale as $u^2$ for low velocities as seen from Equation (\ref{tdec_mark}). This quadratic behavior of the internal dynamics is in agreement with the results found in \citen{buhmann_spectroscopic}, where among other aspects of the internal dynamics of an atom, the decay rate - which is proportional to the Markovian limit of $D(v,t)$ - is found to scale as $u^2$. So far, we have derived a timescale that provides concrete evidence of the motion-induced decoherence on the internal degrees of freedom of the neutral particle. Equation (\ref{tdec_mark}) provides important information on the decoherence process: when the velocity $u$ of the particle is negligible, decoherence will be mainly due to the presence of the vacuum dressed by the dielectric material. When the velocity is increased, the decoherence effects are enhanced. 
Consequently, if we manifestly define decoherence time in this regime as $\tau_{\rm D}\sim a - b\,u^2$, it becomes instructive to study the factor $b/a$ as a relative rating between these two contributions.
We are defining $\tau_{\rm D}$ as the net effect of the environment on the particle while $\tau_{\rm D}|_{u=0}$ is the decoherence time when the particle is static. If velocity effects are insignificant $\tau_{\rm D}/\tau_{\rm D}|_{u=0} \sim 1$. Then, by inspecting the quantity $(\tau_{\rm D}/\tau_{\rm D}|_{u=0} -1)$ we can explore the ratio $b/a$. 
In \textbf{Figure \ref{fig4}}, we show this quantity as a function of velocity for two different frequencies $\tilde{\Delta}$ (energy gap of the two-level  system). Therein, we can easily note a quadratic behavior. The results show that the environmental induced effects and the motion induced effects are strongly dependent on the parameters of the composite system. We must recall that these parameters are introduced by the material of the half-space, the level-spacing of the particle and its velocity. It can be seen in \textbf{Figure \ref{fig4}} that considering an NV-center sliding over an n-doped silicon (n-Si) surface, the net environmentally-induced effects are stronger than those obtained when considering a Rb atom moving over the same surface at the same dimensionless velocity.

\begin{figure}
\centering
 \includegraphics[width=0.9\columnwidth]{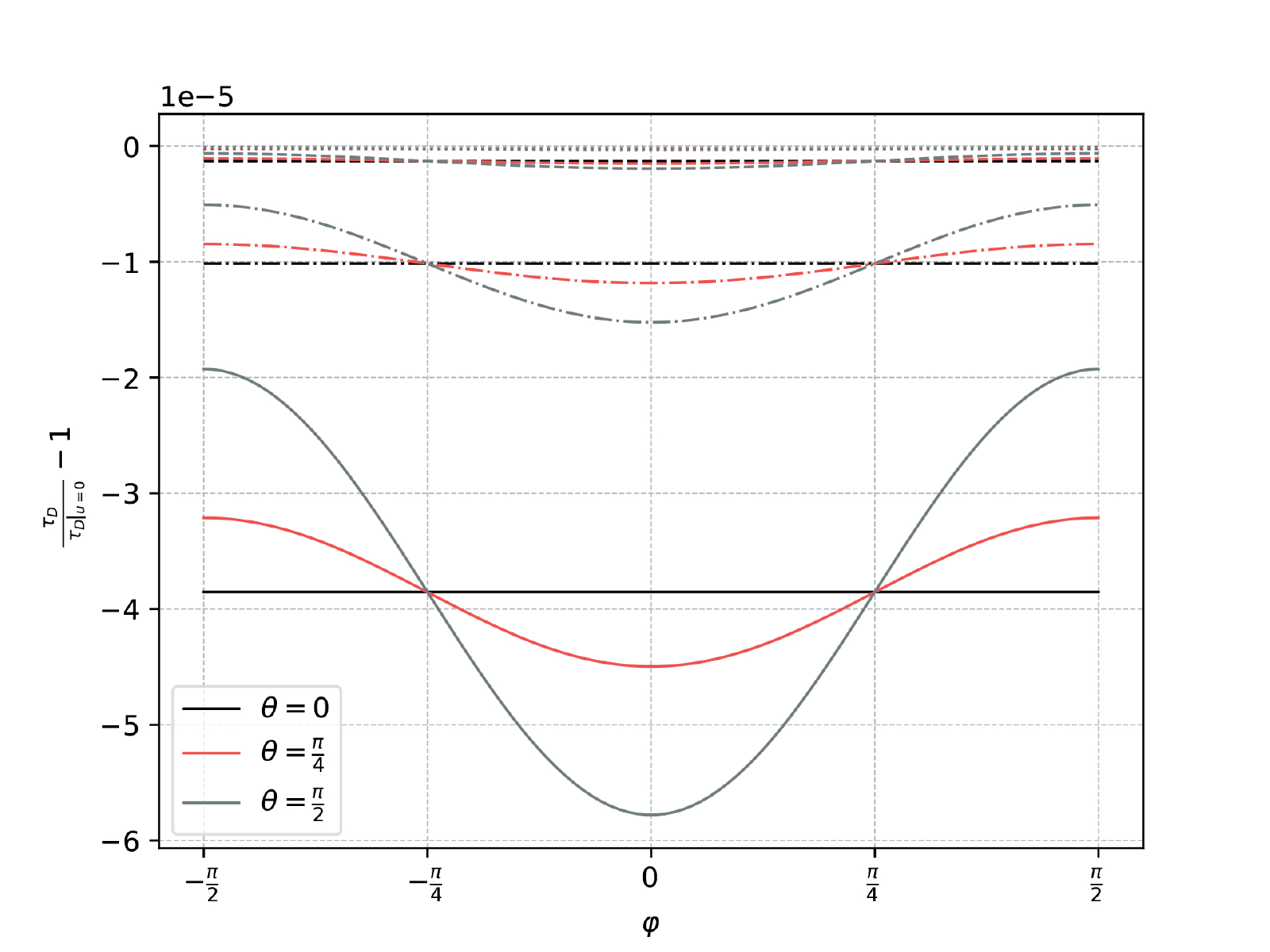}
 \caption{Decoherence time as a function of the polarization direction of the system, which  can be written as $\mathbf{d}=d(\sin(\theta)\cos(\phi)\hat{x} + \sin(\theta)\sin(\phi)\hat{y} + \cos(\theta)\hat{z})$ for a set of combinations of particles and materials. Different line styles represent different system-material combinations in this correspondence: Dotted lines correspond to Rb atom-Au, dashed lines correspond to Rb atom - n-Si, dot-dashed lines to NV center-Au and solid lines correspond to NV center - n-Si.}
 \label{fig5}
\end{figure}

So far, we have studied how the environmentally induced effects on the degrees of freedom of the particle depend upon the level-spacing of the particle and its velocity. However, we still need an insight on how the orientation of the polarization of the dipole $\mathbf{d}=d(\sin(\theta)\cos(\varphi)\hat{x} + \sin(\theta)\sin(\varphi)\hat{y} + \cos(\theta)\hat{z})$ (where $\varphi$ and $\theta$ are the spherical azimuthal and polar angles, respectively) modifies or affects the decoherence process. In \textbf{Figure \ref{fig5}}, we show the $\varphi$ dependence for different fixed $\theta$ values for a set of combinations of particles and surfaces. For the dielectric material we shall consider metals like a gold surface (Au) or an n-doped silicon material (n-Si). Gold has Drude-Lorentz model parameters $\omega_{\text{s}}^{\text{Au}}\sim 9.7 \times 10^{15}$ rad$\,\text{s}^{-1}$ and $\Gamma/\omega_{\text{s}}\sim 0.003$, while the corresponding parameters for n-Si are 
$\Gamma/\omega_{\text{s}}\sim 1$ and $\omega_{\text{s}}^{\text{n-Si}}\sim 2.47 \times 10^{14}$ rad$\,\text{s}^{-1}$. As for the particles (atoms), we shall consider a Rb atom or a single NV center in diamond as an effective two-level system. The different line styles in \textbf{Figure \ref{fig5}} represent different system-material combinations according to: dotted lines represent the Rb atom-Au, dashed lines the Rb atom-n-Si, dot-dashed lines NV center-Au and solid lines correspond to NV center-n-Si. In all cases, the decoherence time has its smallest value when the polarization is perpendicular to the metallic surface ($\theta=0$). If the dipole moment is tilted, the coherence falls sooner when the polarization is in the direction of the velocity. This result is consistent with that reported in \citen{dalvit_nonmarkovianity}, where the authors observed the opposite dependence upon the polarization for quantum friction, by a different
approach to the one performed herein. As for our results, $\tau_{\text{D}}$ decreases implying that decoherence effects are stronger when frictional force is. This constitutes a direct link between decoherence and quantum friction since they exhibit a qualitative inverse proportionality: the larger the decoherence effect (shorter decoherence time), the bigger the frictional force.
The results obtained reinforce the idea that the velocity-dependent effects induced on the atom depend on the material and particle. $\tau_{\text{D}}/\tau_{\text{D}_{u=0}}$ can be enhanced up to a factor $10^2$ by considering an NV center moving over an n-Si coated surface, when compared to a Rb atom moving over a gold-coated surface. 

\vspace{0.3cm}
\subsection{Geometric phase as a quantum sensor}
\label{GP}

Since the work of Berry \cite{Berry}, the notion of geometric phases (GPs) was shown to have important consequences for quantum systems. Berry demonstrated that quantum systems could acquire phases that are geometric in nature. He showed that, besides the usual dynamical phase, an additional phase that was related to the geometry of the space state was generated during an adiabatic evolution, i.e. it depends only on the system's path in parameter space, particularly the flux of some gauge field enclosed by that path. It is also well known that the state of a point-like quantum system with discrete energy levels (such as a two-level atom) acquires a GP when interacting with a quantum field. This phase is dependent on the state of the field. For pure field states, the GP is said to encode information about the number of particles in the field \cite{guridi1}, for initial squeezed states, the phase also depends on the squeezing strength \cite{guridi2}. If the field is in a thermal state, the GP encodes information about its temperature, and so it has been used in a proposal to measure the Unruh effect at low accelerations. It has further been proposed as a high-precision thermometer by considering the atomic interference of two atoms interacting with a known hot source and an unknown temperature cold cavity \cite{martinez}. As the existing bibliography reflects, GPs have become a fruitful venue of investigation to infer features of a quantum system due to their topological properties and close connection with gauge theories of quantum fields. Under suitable conditions, the corrections induced by the presence of the environment can be measured by means of an interferometric experiment (atomic interference) \cite{zeillinger, leek, maclaurin,wood} or by NMR techniques \cite{cucchietti}.
In the case of open quantum evolution, the geometric phase that the system acquires $\phi_g$ differs from that acquired when the evolution is closed $\phi_u$ \cite{tong} since it is now affected by non-unitary effects such as decoherence and dissipation.
This means, in a general case, the phase is $\phi_g = \phi_u + \delta \phi$, where $\delta \phi$ is the correction to the unitary phase induced by the presence of the environment \cite{lombardo1,lombardo2,lombardo3,villar1,villar2,villar3}. 

\vspace{0.3cm}
 A proper generalization of the geometric phase for unitary evolution to a non-unitary evolution is crucial for practical implementations of geometric quantum computation. In \citen{tong}, a quantum kinematic approach was proposed and the geometric phase for a mixed state under non-unitary evolution has been defined as
 
\begin{align} \nonumber
\phi_{\rm g} = {\rm arg}\left\lbrace\sum_k \sqrt{ \varepsilon_k (0) \varepsilon_k (\tau)} \langle\Psi_k(0)|\Psi_k(\tau)\rangle \right.\\ \left.\times e^{-\int_0^{\tau} dt \langle\Psi_k| \frac{\partial}{\partial t}| {\Psi_k}\rangle}\right\rbrace , \label{fasegeo}
\end{align}

where $\varepsilon_k(t)$ are the eigenvalues and $|\Psi_k\rangle$ the eigenstates of the reduced density matrix $\rho_{\rm s}$ (solution of Equation (\ref{meq_exacta})). In Equation \ref{fasegeo}, $\tau$ denotes the amount of time it would take for the system to complete a cyclic evolution if it were isolated from the environment. It has been argued that the observation of GPs should be done in times long enough to obey the adiabatic approximation but short enough to prevent decoherence from deleting all phase information. This means that while there are dissipative and diffusive effects inducing corrections to the unitary GP, the system should maintain its purity for several cycles, allowing the GP to be observed. It is important to note that if the effects of noise induced on the system are of considerable magnitude, the coherence terms of the quantum system are rapidly destroyed and the GP literally disappears \cite{lombardo1}. The conclusions that can be reached from the solution of the master equation derived in the above section render a good scenario for measurements of the GP as compared to other models. By use of Equation (\ref{fasegeo}), we can compute the geometric phase accumulated by the best combination of particle and material obtained above: a NV-center traveling above a n-Si coated surface. We expect the geometric phase to have two differently induced contributions: one originated in the dressed quantum field $\delta \phi_{{\bf u}=0}$ and another originated in the velocity of the atom $\delta \phi_{{\bf u}\neq0}$. 

\vspace{.3cm}
In the Markovian limit, the $D$ and $\zeta$ kernels can be written as $ D(v,t)= \zeta(v,t) \sim \xi$ with 

\begin{align}\nonumber
\xi =& \frac{r_0}{4}\;\left(d^{(i)} + \frac{3}{8}d^{(a)}\, u^2\,\partial^2_{\Tilde{\Delta}}\right)\frac{\Tilde{\Delta}\Tilde{\Gamma}}{\left(\Tilde{\Delta}^2-1\right)^2+\Tilde{\Delta}^2\Tilde{\Gamma}^2} \\
 =& \xi^{(0)} + \xi^{(2)}\;u^2.
\end{align}

By performing the same expansion on the dissipation kernel $f = f^{(0)} + f^{(2)}\;u^2 $ and further applying the secular approximation (also known as post-trace RWA) we can obtain an expression for the geometric phase

\begin{widetext}
\begin{align*}
 \phi_{\text{g}} =& \arg\left[\sin(\alpha)\cos^2(\alpha)e^{-2\xi \tau+\mathfrak{i}(2f+\Tilde{\Delta})\tau}-\frac{\sin(\alpha)}{2}\left(1 -2\sin^2(\alpha)e^{-4\xi \tau} - \sqrt{1 + 4\sin^4(\alpha)(e^{-8\xi \tau} -e^{-4\xi \tau}) }\right)\right] - \\[1em]
 &-\int^\tau_0 dt \frac{2\sin^2(\alpha)\cos^2(\alpha)e^{-4\xi t}}{ 1 + 4\sin^4(\alpha)(e^{-8\xi t} -e^{-4\xi t}) - (1 -2\sin^2(\alpha)e^{-4\xi t})\sqrt{1 + 4\sin^4(\alpha)(e^{-8\xi t} -e^{-4\xi t})}}(2f+\Tilde{\Delta}).
\end{align*}
\end{widetext}

\vspace{.3cm}

\begin{figure}[H]
\centering
 \includegraphics[width=0.9\columnwidth]{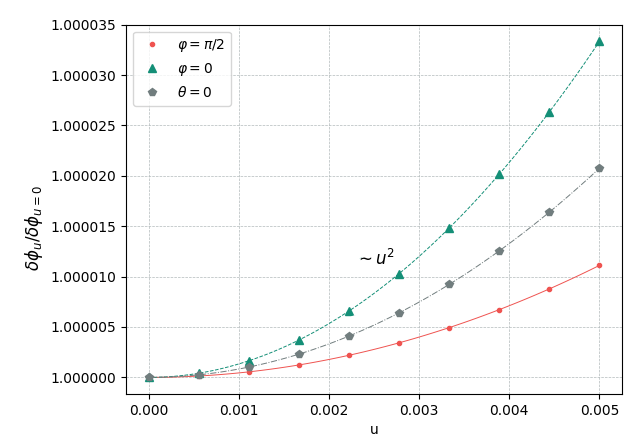}
 \caption{Correction to the geometric phase  $\delta \phi_u$ (normalized with respect to the static correction $\delta \phi_{u=0}$) as a function of the dimensionless velocity $u$, considering different polarization directions of the particle. Red line (circle) corresponds to $\theta = \pi/2$ and $\varphi = \pi/2$ values which yields a polarization along the $y$ axis, parallel to the surface and perpendicular to the motion. Gray line with (hexagon) corresponds to the $\theta =0$ case, yielding a polarization perpendicular to the dielectric surface, along the $z$ axis. Green line with (triangle) corresponds to the case in which the atom is polarized in the direction of motion $\theta = \pi/2$ and $\varphi = 0$. As it has been reported, for very small velocities the presence of the quantum field dressed by the dielectric is dominant. However, for bigger velocities, the theory predicts that $\delta \phi_{u\neq0}$ becomes relevant making it possible to detect differences as the geometric phase accumulates. Parameters used: $N = 100$, $\tilde{\Gamma}=1$, $r_0/\omega_{\text{s}}=10^{-2}$, $\tilde{\Delta}=0.2$. }
 \label{fig6}
\end{figure}

 We show in \textbf{Figure \ref{fig6}} the accumulated environmentally induced correction  $\delta \phi$ over $N=100$ natural cycles. We have normalized the correction by the environmentally induced correction when the particle is static $\delta \phi_{u=0}$. If the correction to the geometric phase induced by the motion is not relevant, $\delta \phi/\delta \phi_{u=0}$ will remain close to unity, an effect observed for very small dimensionless velocities. However, when the value of $u$ is substantial, the correction due to the velocity becomes more important and the GP begins to show different behaviors as the dipole orientation is varied. The correction acquired, furthermore, can be fitted by a $u^2$ behavior.

\vspace{0.3cm}
All in all, we can state that the motion of the particle induces a phase-lag on the fluctuating dipole that results in dynamical effects. The corrections due to the quantum fluctuations are evidenced as stochastic variations in the energy gap of the two-level system. This random variation produces noise effects on the internal degrees of freedom that induces corrections in the GP of the two-level system.  Recalling the results obtained so far, we can assure this model is a good scenario for the measurement of the GP and its correction.
If the correction induced by the velocity becomes relevant enough, it yields the opportunity to detect traces of the velocity in the correction of the geometric phase. Taking into account the experimental values used in current measurements and the free parameters of the numerical simulations of the model, we can show that it is possible to detect a velocity dependence in the corrections of the geometric phase, as it accumulates in time.

\begin{figure}[h]
 \centering
 \includegraphics[width=.95\linewidth]{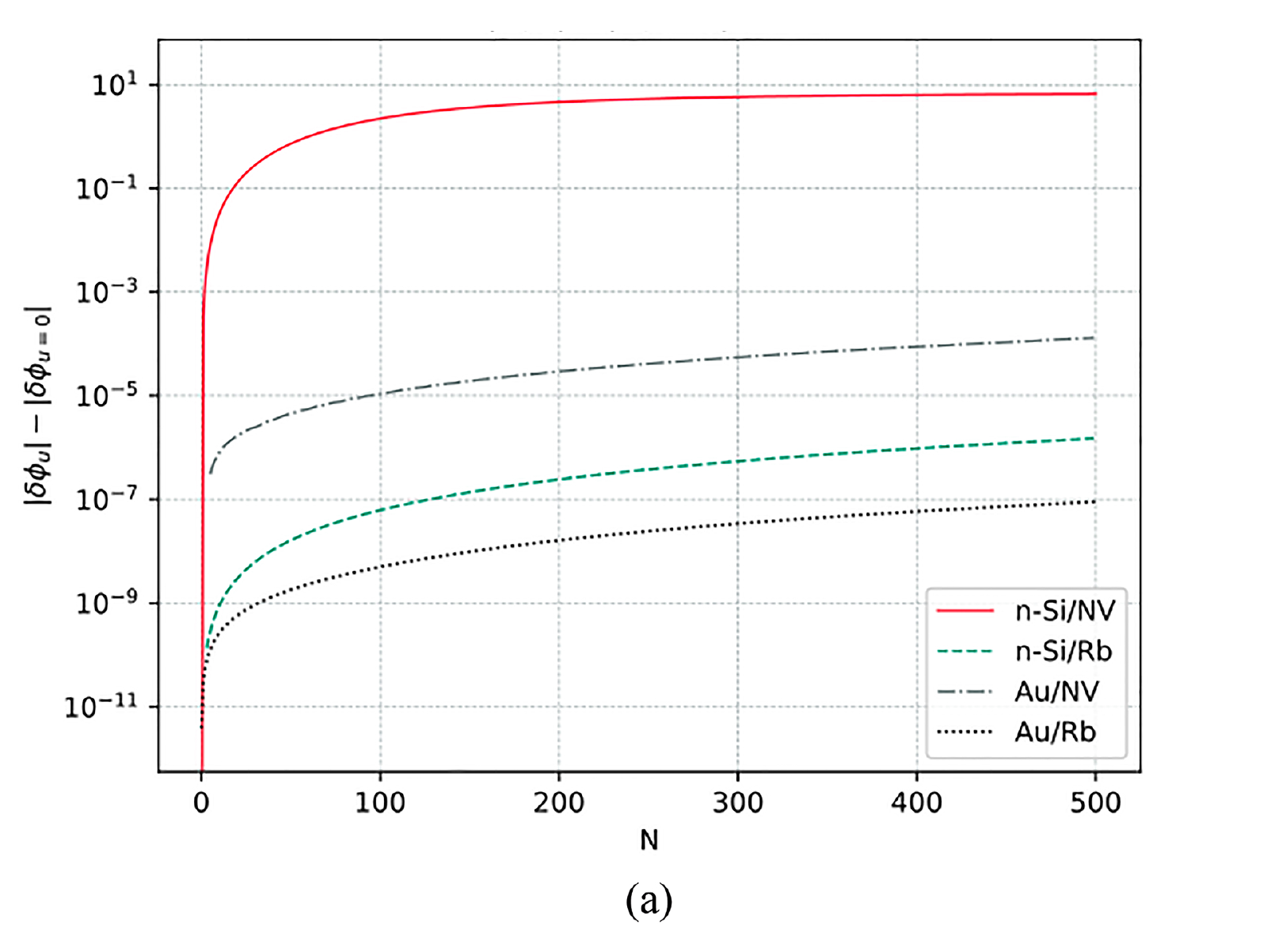}
 \includegraphics[width=.95\linewidth]{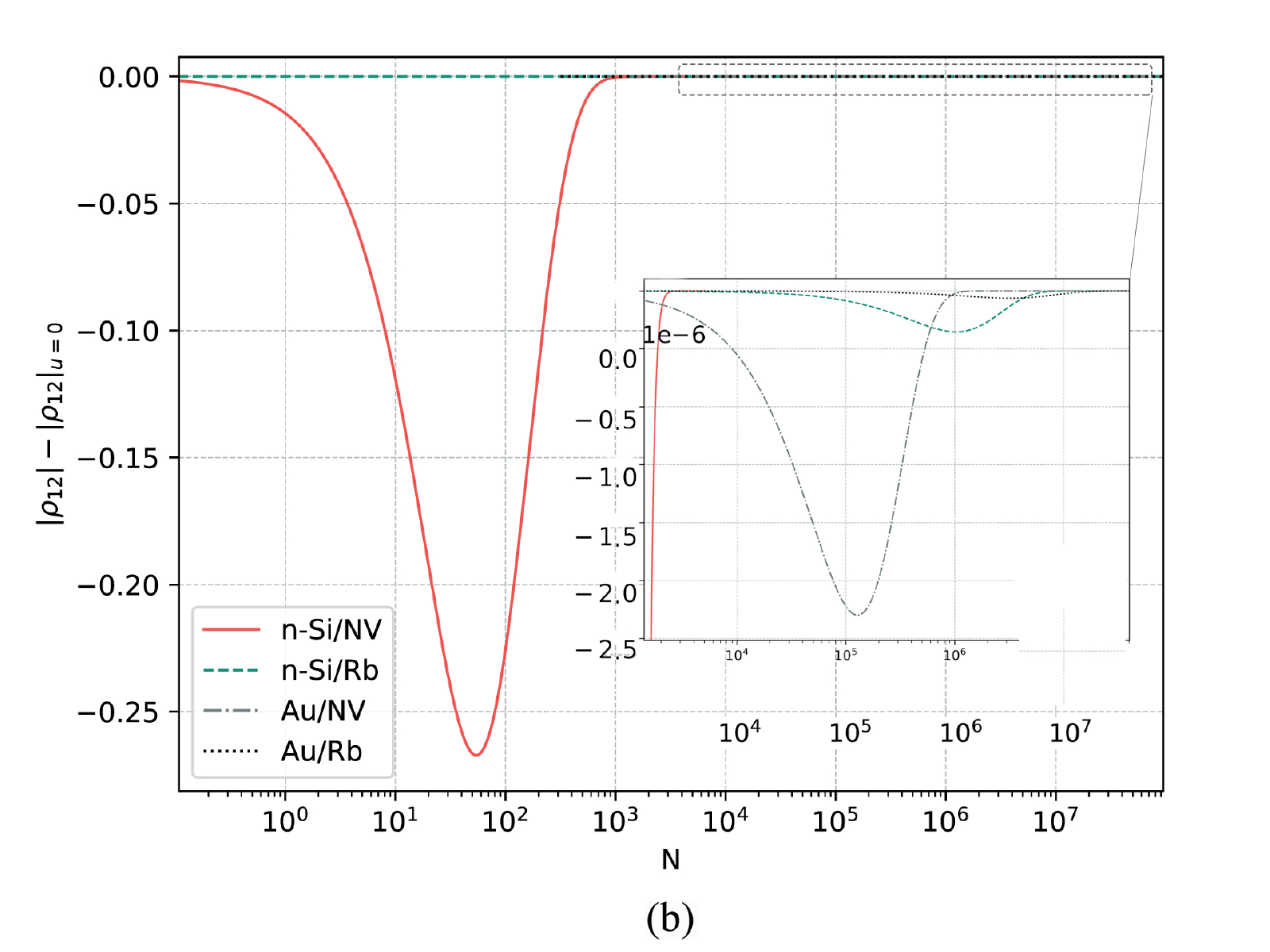}
 \caption{Velocity dependent correction to the geometric phase $\delta \phi_{u\neq0} = \delta \phi - \delta \phi_{u=0}$ evolution (a) and difference in the coherence as cycles proceed (b), both for different systems and different materials. Reference: Different line styles represent different system-material combinations in this correspondence: Black dotted lines correspond to Rb atom-Au, green dashed lines correspond to Rb atom - n-Si, gray dot-dashed lines to NV center-Au and red solid lines correspond to an NV center of frequency $\tilde{\Delta}=4\times10^{-4}$, moving over an n-Si surface. For this last combination, an $N=100$ range corresponds to 6.36 ns and the adimensional velocity $u = 1.5 \times 10^{-4}$}
 \label{fig9}
\end{figure}

\vspace{0.3cm}
\textbf{Figure \ref{fig9} (a)} shows the velocity dependent correction to the geometric phase ($\delta \phi_{u\neq0}$) for different particles moving over an n-Si or an Au surface. 
All combinations exhibit the same qualitative behavior. However, the timescale and magnitude of the induced effects are considerably different. We can note that, once again, the NV center on an n-Si-coated surface is the combination that presents the most promising results, accumulating a geometric phase of order unity over N=500 steps (corresponding to a temporal range of approximately 30 ns for an NV center with adimensional level spacing $\tilde{\Delta}=4\times10^{-4}$), which is not only experimentally detectable with current technologies but also several orders of magnitude greater than the accumulated phase acquired in N = 500 steps by a rubidium atom in identical conditions. It is important to remark that the greater the velocity achieved, the earlier the velocity-dependent corrections become relevant.

The same hierarchy of effects is seen in \textbf{Figure \ref{fig9} (b)}, where the evolution of the difference in the coherence is shown for various particle-material combinations. Once again, all four studied combinations exhibit the same qualitative behavior while differing in magnitude and timescale. As observed in the geometric phase case, the velocity-induced effect, while still small, is considerably larger for an NV center moving over an n-Si surface than for any other considered combination.

\vspace{0.3cm}
\section{Experimental section}
\label{experiment}
\begin{figure}[h]
\centering
 \includegraphics[width=0.85\linewidth]{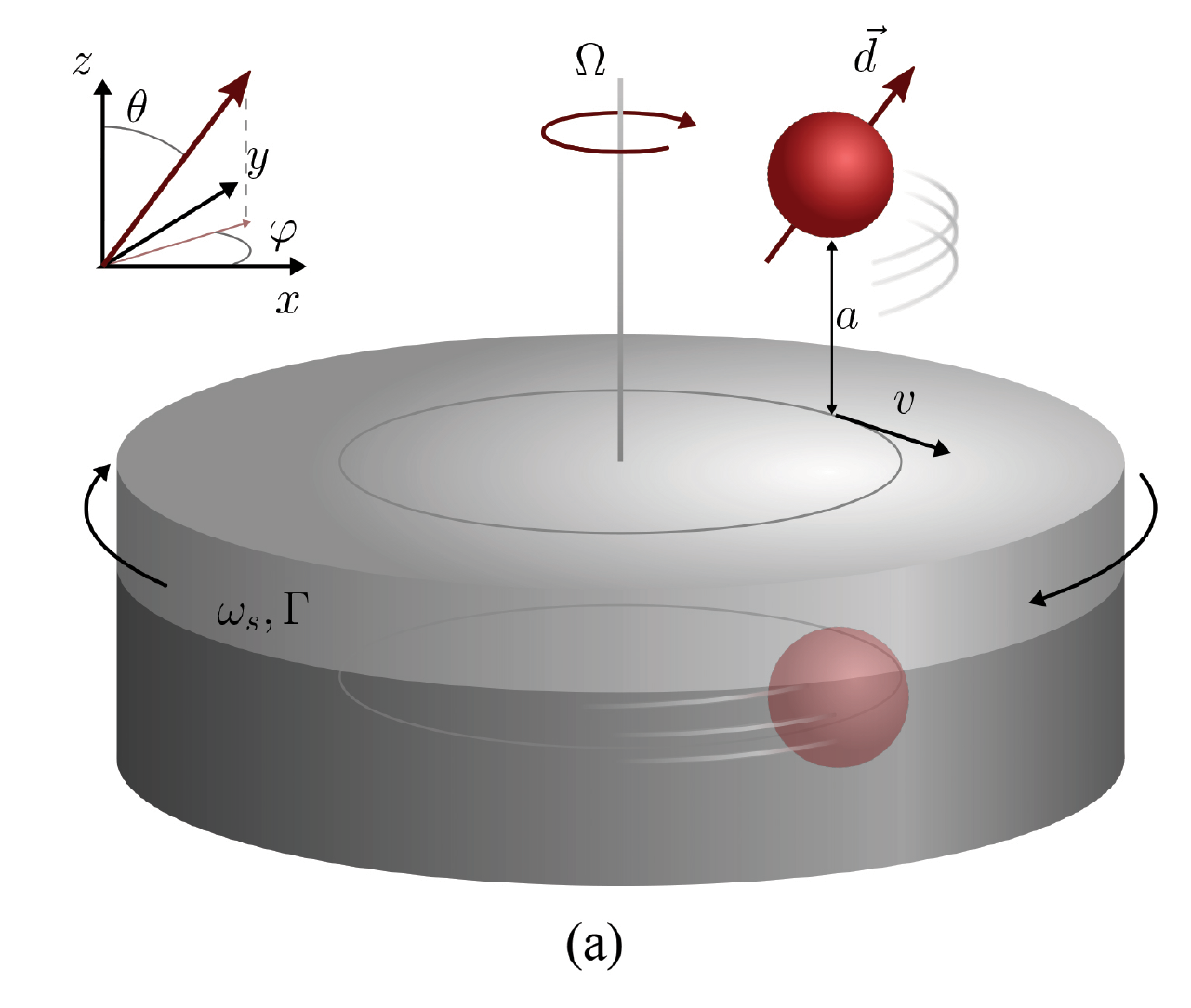}
 \includegraphics[width=0.85\linewidth]{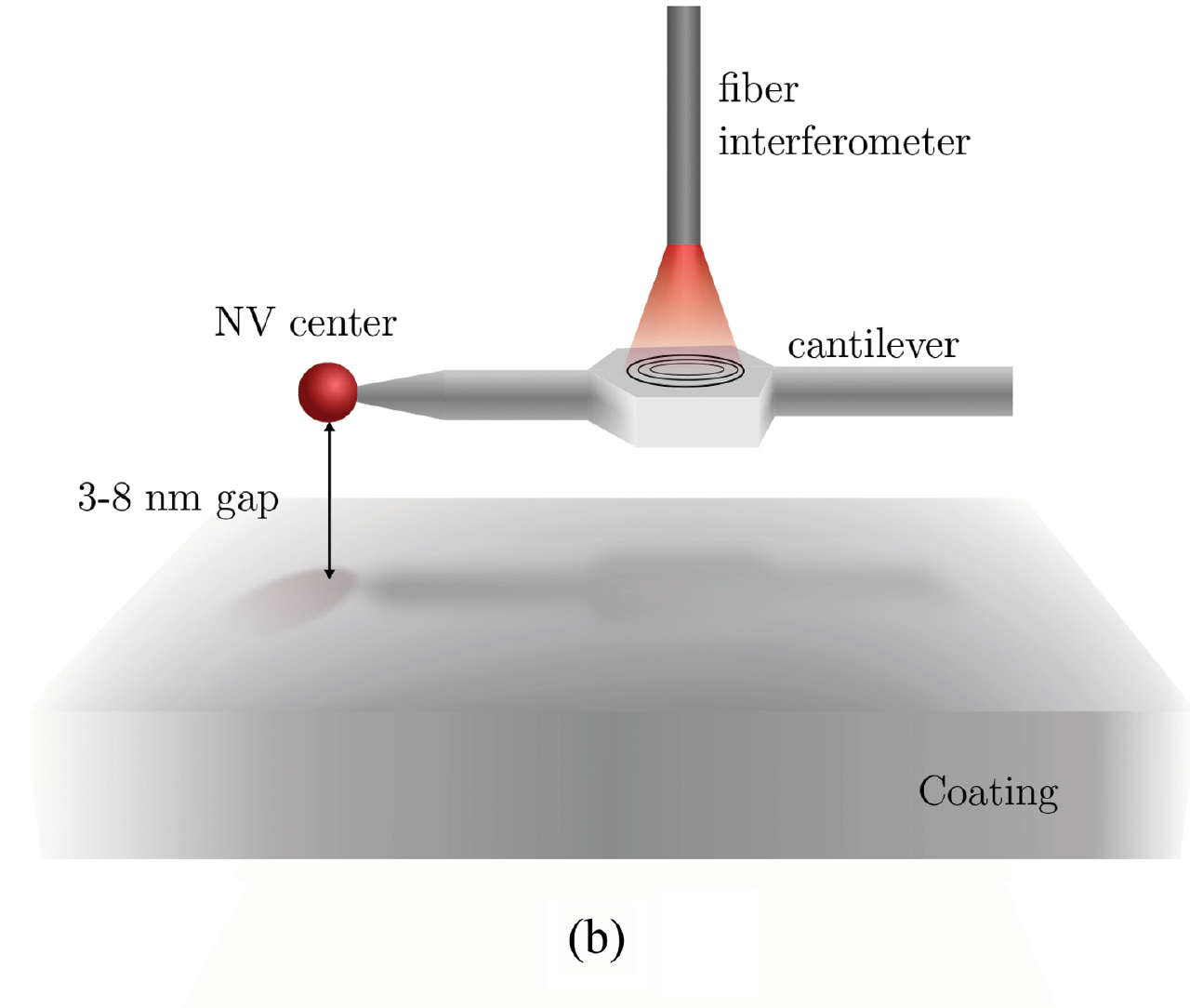}
 \caption{(a) Scheme of the proposed experimental setup. An Au-coated Si disk rotates at angular velocity $\Omega$. The diamond NV center is placed at a distance $a$. (b)  Scheme of the setup. A diamond with an NV center is placed at the end of an AFM system. The AFM is used to keep the gap $a$ constant within 1 nm.}
 \label{fig7}
\end{figure}

A feasible experimental setup to measure the correction induced on the accumulated geometric phase would be based on the use of an NV center in diamond as an effective two-level system at the tip of a modified atomic force microscope (AFM). The distance can be controlled from a few nanometers to tenths of nanometers with subnanometer resolution. The NV center consists of a vacancy, or missing carbon atom in the diamond lattice, lying next to a nitrogen atom which has substituted one of the nearest neighbours of the vacancy. The electron spin is the canonical quantum system and the NV center offers a system in which a single spin can be initialized, coherently controlled, and measured. It is also possible to mechanically move the NV center. The NV system is chosen as it presents itself as an excellent tool for studying geometric phases \cite{nvgp}. 

\vspace{0.3cm}
The dielectric surface is made of an n-doped Si wafer where the doping density is adjusted to maximize the effects, as depicted in {\bf Figure 4}. 
This n-Si-coated Si disk is mounted on a turntable as schematically indicated in \textbf{Figure \ref{fig7}}. Although we are using a rotating table, non-inertial effects can be completely neglected to model a particle moving at a constant speed on the material sheet. Since it is critical to keep the separation uniform to prevent spurious decoherence, it is important to assess the plausibility of the proposed experimental setup. We have checked that a $12$~cm diameter Si disk can be rotated up to $\Omega= 2\pi \times 7000$ Hz. This yields tangential velocities up to $\sim$~ 5300~ms$^{-1}$, indicating that the velocities used in {\bf Figure \ref{fig9}} of 30~ms$^{-1}$ (corresponding to $u=v/(\omega_s a)=1.5\times 10^{-4}$) are readily accessible. Within the whole range of speeds investigated, the measured wobble of the turntable is of the order of $10^{-8}$ radians (i.e. the vertical motion is 1~nm at the edge of the disk). The experiment is doable at $a=10$~nm, with $\delta a$ (possible fluctuations in distance) induced decoherence effects being negligible compared to the quantum friction ones \cite{farias_nature}. 


\vspace{0.3cm}
Using this experimental setup, the geometric phase can be computed in a tomographic manner \cite{cucchietti} by measuring the elements of the reduced density matrix of the system and appealing to the definition of the GP for its computation.
These elements have been studied under the assumption that values of $a$ ranged between 3 and 10 nm (as the one experimentally measured). As we have shown, the magnitude of the motion-induced correction for an NV-center traveling at a constant velocity over an n-Si coated surface yields traces of quantum friction that can be detected with current technologies. The NV level structure and the techniques for measurement and manipulation of the electronic spin are well documented, see for example \citen{WrachtrupPRL, WrachtrupSci,WrachtrupPR}.
Basically, the electronic ground state ($|g\rangle$) is a spin triplet with a splitting $\Delta \sim 2\pi \times 2.88$~GHz between the $|m_s=0\rangle$ and the $|\pm 1\rangle$ spin states. Off-resonant excitation pumps the NV into the excited state $| e\rangle$ which has a lifetime of a few ns, while taking some of the $|\pm 1\rangle$ states to the metastable singlet $|s\rangle$, with a lifetime $\sim 300$ ns. The state $|s\rangle$ preferentially decays to the $|m_s=0\rangle$ state, providing a spin-dependent fluorescence rate. The spin of the NV can be manipulated using resonant microwave pulses.
The $|\pm1\rangle$ degeneracy is lifted by a magnetic field, and then the system is effectively converted into a two-level system.

\vspace{0.3cm}
The effects of quantum friction on the two-level system can be observed by doing experiments at zero and finite relative velocities between the NV center and the n-doped Si system. These effects can be detected, in principle, both in the decoherence times and in the accumulated geometric phase. Decoherence can be measured by applying a $\pi$ pulse to the $|0\rangle$ state and measuring the fluorescence intensity as a function of a delay time $\tau$. As seen in {\bf Figure~\ref{fig9}} the differences in coherent lifetimes are small and impossible to detect after N = 200 steps, corresponding to about 15~ns for the proposed sample. On the other hand, as {\bf Figure~\ref{fig9}} indicates, the effects are much more pronounced in the accumulated geometric phase, such that after a relatively short time of less than $1~ \mu$s the phase difference accumulates to a value of $\sim 2 \pi$ which can be readily measured. With the parameters used in the calculation, it appears geometrical dephasing effects can be measured by applying a single Ramsey sequence. Starting with the system at $|m_s=0\rangle$, a $\pi/2$-pulse would leave it at $|\Psi\rangle =|0\rangle+e^{i\phi}|+1\rangle$ and, left in this state for a time $\tau$, $|+1\rangle$ acquires a relative phase which is induced by quantum friction. The application of a second $\pi$/2-pulse will convert the phase into a population difference between the two spin states, which can be measured by fluorescence. If the simple Ramsey sequence is not sufficient to observe the phase difference, usual spin-echo techniques can be used. A spin-echo sequence, comprised of a Ramsey sequence of length 2$\tau$ with a $\pi$-pulse applied after free precession time $\tau$, will cancel out entirely the dephasing effect arising from background contributions as those arising from $^{13}$C environment. During the first $\tau$ of free
precession the detuning $\delta$ of the microwave signal from resonance will induce a relative phase $\delta\phi = \delta \times \tau$. The $\pi$-pulse changes the relative phase to $-\delta\phi$ and the second part of the free precession will produce a null net effect on the bath induced dephasing.


\section{Conclusion}
\label{conclusions}

In this article we have studied the complete dynamics of a two-level system in motion relative to a semi-infinite metallic material in the electromagnetic field vacuum, and characterized the effects of motion in the dynamics as an alternative to the explicit computation of QF.
By obtaining the reduced density matrix, we have estimated the decoherence timescale at which coherence is are strongly suppressed.
We have further obtained an analytical expression for the decoherence time in the secular regime and Markovian limit. With this expression an additional comparison among the parameters introduced by the model which have a strong influence on the system's dynamics could be made. Through analytic considerations we have shown how both, the net effect of the composite environment on the particle and the velocity-induced effect, are strongly dependent on the material parameters and the system level spacing, allowing  the magnitude of the effects to be
strengthened or weakened by a sensible choice of systems. 
We have found results for the decoherence time in agreement with those existing in the literature for quantum friction, showing that a qualitative inverse proportionality relates them. Thus, a link between the decoherence time and the quantum frictional force can be established since non-contact friction seems to enhance the decoherence of the moving atom and suggests that measuring decoherence times could be used to indirectly demonstrate the existence of quantum friction. 

\vspace{0.3cm}
We have studied the correction to the unitary geometric phase and realized it can be decomposed into different contributions: on the one hand, a correction induced by the field vacuum dressed by the presence of a dielectric sheet and, on the other hand, a correction induced by the velocity of the particle which is moving in front of the dielectric sheet. We have also shown that after many cycles, the correction to the accumulated GP due to the velocity of the particle becomes relevant. In this context, we have proposed an experimental setup that determines the feasibility of an experiment that would be the first to track traces of quantum friction through the GP acquired by a two-level system. It is important to remark that the mere presence of a velocity contribution in the noise corrections to the phase is an indication of the frictional effect over the quantum degrees of freedom of the particle.
All in all, we have found a scenario to indirectly detect QF by measuring the GP accumulated by a particle moving above of a dielectric plate.
The emerging micro and nanomechanical systems promising new applications in sensors and information technology may suffer or benefit from non-contact quantum friction. For these applications, a better understanding of non-contact friction is needed. It’s detection will be a step forward.

\vspace{0.3cm}
\medskip
\textbf{Acknowledgements} \par 

The work of F. C. L., P. I. V. and L. V. was supported by ANPCyT 
through grants PICT-2018-3801, CONICET, and Universidad de Buenos Aires through grants UBACyT 2018; Argentina.
R. S. D. acknowledges support from the National Science Foundation through grants PHY-1607360 and PHY-1707985 and financial and technical support from the IUPUI Nanoscale Imaging Center, the IUPUI Integrated Nanosystems Development Institute, and the Indiana University Center for Space Symmetries. P.I.V acknowledges ICTP-Trieste Associate Program.

\medskip

\bibliography{apsrev4-1}

{}

\end{document}